\documentclass[fleqn,usenatbib]{mnras}

\usepackage{newtxtext,newtxmath}
\usepackage[T1]{fontenc}

\DeclareRobustCommand{\VAN}[3]{#2}
\let\VANthebibliography\thebibliography
\def\thebibliography{\DeclareRobustCommand{\VAN}[3]{##3}\VANthebibliography}

\usepackage{graphicx}
\usepackage{amsmath}
\usepackage{subcaption}

\title[Light-curve Modelling of RETs Powered by Outflow]{Light-curve Modelling for The Initial Rising Phase of Rapidly-evolving Transients Powered by Continuous Outflow}

\author[K. Uno \& K. Maeda]{
Kohki Uno$^{1}$\thanks{E-mail: k.uno@kusastro.kyoto-u.ac.jp}
and Keiichi Maeda$^{1}$
\\
$^{1}$Department of Astronomy, Kyoto University, Kitashirakawa-Oiwake-cho, Sakyo-ku, Kyoto, 606-8502, Japan
}

\date{Accepted 2023 March 15. Received 2023 March 13; in original form 2022 November 11}

\pubyear{2022}

\begin{document}

\label{firstpage}
\pagerange{\pageref{firstpage}--\pageref{lastpage}}
\maketitle

\begin{abstract}

A wind-driven model is a new framework to model observational properties of transients that are powered by continuous outflow from a central system. While it has been applied to Fast Blue Optical Transients (FBOTs), the applicability has been limited to post-peak behaviours due to the steady-state assumptions; non-steady-state physics, e.g., expanding outflow, is important to model the initial rising phase. In this paper, we construct a time-dependent wind-driven model, which can take into account the expanding outflow and the time evolution of the outflow rate. We apply the model to a sample of well-observed FBOTs. FBOTs require high outflow rates ($\sim 30$\,M$_{\sun}$\,yr$^{-1}$) and fast velocity ($\sim 0.2-0.3c$), with the typical ejecta mass and energy budget of $\sim 0.2$\,M$_{\sun}$ and $\sim 10^{52}$\,erg, respectively. The energetic outflow supports the idea that the central engine of FBOTs may be related to a relativistic object, e.g., a black hole. The initial photospheric temperature is $10^{5-6}$\,K, which suggests that FBOTs will show UV or X-ray flash similar to supernova shock breakouts. We discuss future prospects of surveys and follow-up observations of FBOTs in the UV bands. FBOTs are brighter in the UV bands than in the optical bands, and the timescale is a bit longer than in optical wavelengths. We suggest that UV telescopes with a wide field of view can play a key role in discovering FBOTs and characterizing their natures.

\end{abstract}


\section{Introduction} \label{sec:intro}

Thanks to new-generation optical surveys, e.g., Pan-STARRS \citep[Panoramic Survey Telescope And Rapid Response System;][]{Kaiser2002SPIE}, ATLAS \citep[Asteroid Terrestrial-impact Last Alert System;][]{Tonry2018PASP} and ZTF \citep[Zwicky Transient Facility; e.g.,][]{Kulkarni2018ATel}, the number of discoveries of rapidly-evolving transients \citep[RETs;][]{Ofek2010ApJ,Poznanski2010Science,Kasliwal2010ApJ}, which show short timescales of $\lesssim 10$\,days, has been dramatically increasing in the latest decade \citep{Drout2014ApJ,Tanaka2016ApJ,Pursiainen2018MNRAS,Tampo2020ApJ,Ho2021arXiv}. Furthermore, the launch of the Rubin Observatory \citep{LSST2009arXiv} is expected to advance the transient research to the next stage. 

The RETs are roughly classified into three populations: Type IIb/Ib Supernovae (SNe), Type IIn/Ibn SNe, and AT2018cow-like transients, i.e., Fast Blue Optical Transients \citep[FBOTs;][]{Ho2021arXiv}. The nature of FBOTs is, in particular, unclear since the sample of well-observed FBOTs is very limited; AT2018cow \citep{Prentice2018ApJ,Perley2019MNRAS}, AT2018lug \citep{Ho2020ApJ}, AT2020xnd \citep{Perley2021MNRAS}, AT2020mrf \citep{Yao2021arXiv}, CSS161010 \citep{Coppejans2020ApJ}, and MUSSES2020J \citep{Jiang2022ApJ}. FBOTs exhibit high optical peak luminosity over $-20$\,mag with blue continuum spectra dominated by high-temperature blackbody radiation of $\gtrsim 30000$\,K \citep[e.g.,][]{Perley2019MNRAS, Ho2020ApJ}. The rising timescale is less than $\sim 5$\,days \citep{Ho2021arXiv}, which is much faster than the typical timescale of other RETs. The rapidly rising luminosity to the peak indicates that FBOTs have very fast ejecta over $0.1c$ \citep[e.g.,][]{Perley2019MNRAS,Ho2019ApJ}, where $c$ is the light speed. Furthermore, on top of the optical emission, FBOTs are bright in both X-ray and radio bands \citep[e.g.,][]{Margutti2019ApJ,Ho2019ApJ,Coppejans2020ApJ,Yao2021arXiv,Bright2022ApJ}. 

To explain the peculiar observational properties, some light-curve models have been proposed, including continuous outflow powered by a central engine \citep[e.g.,][]{Piro2020ApJ, Uno2020ApJ}, radiation from a fallback accretion disk around a stellar-mass black hole \citep[BH, e.g.,][]{Kashiyama2015MNRAS, Quataert2019MNRAS, Margutti2019ApJ}, circumstellar interaction \citep[e.g.,][]{Fox2019MNRAS, Leung2020ApJ, Pellegrino2022ApJ}, an electron-capture-induced collapse of an ONeMg white dwarf \citep[e.g.,][]{Lyutikov2019MNRAS,Lyutikov2022arXiv}, mass ejection via a common-envelope phase \citep[e.g.,][]{Soker2019MNRAS, Soker2022RAA, Metzger2022arXiv}, tidal disruption events \citep[TDEs, e.g.,][]{Perley2019MNRAS, Kuin2019MNRAS, Uno2020ApJL}, and a magnetar formation \citep[e.g.,][]{Fang2019ApJ, Mohan2020ApJ}. However, most, if not all, of the proposed models aim to explain only some of the observational features of FBOTs, and thus the origin of FBOTs remains unanswered. We expect that the observational data in the rising phase, which are currently lacking, will play a key role in distinguishing different models. Indeed, MUSSES2020J exhibits characteristic colour evolution toward the peak \citep{Jiang2022ApJ}, which will potentially lead to clarification of the radiation mechanisms of FBOTs. 

High-cadence and wide-field surveys, such as ZTF \citep{Bellm2019PASP}, Tomo-e Gozen survey \citep{Sako2018SPIE}, and Subaru/Hyper Suprime-Cam survey \citep{Tanaka2016ApJ, Jiang2017Nature, Tominaga2019ApJ}, ultimately added by the Rubin observatory \citep{LSST2009arXiv}, will eventually increase the number of FBOT samples with initial rising phase detected. Therefore, it is important to present predictions of light-curve properties by each model before such observational data become available. 

\citet{Uno2020ApJ} proposed a `(steady-state) wind-driven model', which describes expected properties of transients powered by continuous outflow characterized by the outflow rate ($\dot{M}$), the outflow velocity ($v$), and the outflow-launching radius ($R_{\rm eq}$). The steady-state assumption is reasonable for modelling the light curves in the post-peak phases (see, \citet{Uno2020ApJ} and \citet{Uno2020ApJL}). However, it is clear that the steady-state model does not apply to the rising phase, where non-steady-state physics, e.g., the expanding outflow and time evolution of the outflow rate, is important. In order to model the dramatically-evolving initial phase, the model should be expanded to a `time-dependent' wind-driven model \citep[see also][]{Piro2020ApJ}.

In this paper, we propose a time-dependent wind-driven model, which is an updated version of the steady-state model framework. We apply the model to the initial phase of a sample of FBOTs; AT2018cow, AT2018lug, AT2020mrf, and AT2020xnd, and try to constrain the outflow properties of FBOTs through their properties in the initial rising phase. The paper is structured as follows. In Section \ref{sec:2}, we describe an analytical setup of the time-dependent wind-driven model. In Section \ref{sec:3}, we show the general behaviour of the time-dependent model and apply the model to a sample of FBOTs. In Section \ref{sec:4}, we discuss the future observational strategy for FBOTs. Finally, the paper is closed in Section \ref{sec:5} with conclusions.

\section{Time-Dependent Wind-Driven Model} \label{sec:2}

The basic formalism is almost the same as the previous `wind-driven model' \citep{Uno2020ApJ, Uno2020ApJL}, but here, we consider the following two additional constraints: (I) expanding outflow front ($R_{\rm out} (t)$), and (II) time-evolving outflow rate ($\dot{M}(t)$). 
In the present work, we assume a constant outflow velocity ($v$). This assumption is reasonable in the initial rising phase, since key physical properties are mainly controlled by the outflow front. In the initial phase, the ejecta is fully optically thick, and the observational properties are essentially determined solely by the outermost front, regardless of the inner wind properties, including outflow velocity evolution, as we show later. The velocity evolution becomes important only in the later, post-peak phase. This treatment reduces the number of free parameters to model the initial phase as a focus of the present work.

The outflow front expanding with constant velocity is given as follows:
\begin{equation}
    R_{\rm out} (t) = R_{\rm eq} + vt, 
\end{equation}
where $R_{\rm eq}$ is the outflow-launching radius where the equipartition between the internal and kinetic energy densities is assumed. Considering physically-motivated situations, we assume that the outflow rate follows a power-law function, i.e.,
\begin{equation}
    \dot{M}(t) = \dot{M}_{0}\left(1 + \frac{t}{t_{0}} \right)^{\beta},
\end{equation}
where $\dot{M}_{0}$ is the initial outflow rate, $t_{0}$ is the typical timescale for the outflow, and $\beta$ is the power-law index; an accretion-powered outflow corresponds to the index of $\beta = -5/3$, and a radiative outflow has the index of $\beta = -4/3$. Then, the density at a given radius $r$ and time $t$ is defined as follows:
\begin{equation}
    \rho (r, t) = \frac{\dot{M}\left(t - \frac{r - R_{\rm eq}}{v}\right)}{4\pi r^{2} v}.
\end{equation}

Under this configuration, the optical depth for electron scattering ($\tau_{\rm s}$) and the effective optical depth ($\tau_{\rm eff}$; considering not only electron scattering but also absorption processes) are defined as follows;
\begin{equation}
    \tau_{\rm s,eff} (r, t) = \int^{R_{\rm out}(t)}_{r} \kappa_{\rm s,eff} \left(\rho(r, t), T(r, t)\right) \rho(r, t) dr,
\end{equation}
where $\kappa_{\rm s}$ and $\kappa_{\rm eff}$ are the electron-scattering opacity and the effective opacity, respectively, and $T(r, t)$ describes the temperature structure (see below). Here, assuming the solar composition, the electron-scattering opacity is given as $\kappa_{\rm s} = 0.34$\,cm$^{2}$\,g$^{-1}$. The effective opacity is given as follows:
\begin{equation}
  \kappa_{\rm eff} = \sqrt{3(\kappa_{\rm s}+ \kappa_{\rm a})\kappa_{\rm a}} \approx \sqrt{3\kappa_{\rm s} \kappa_{\rm a}}\quad(\kappa_{\rm s} \gg \kappa_{\rm a}).
\end{equation}
Assuming the Kramer's opacity, we use $\kappa_{\rm a} = \kappa_{0}\left(\frac{\rho}{\rm{g\,cm^{-3}}}\right) \left(\frac{T}{\rm K}\right)^{-7/2}$ with $\kappa_0 = 2 \times 10^{24}$\,cm$^{2}$\,g$^{-1}$ \citep{Piro2020ApJ}.

Using the above formalism, we define some typical physical scales; the photon-trapped radius ($R_{\rm ad}$), the colour radius ($R_{\rm c}$), and the photospheric radius ($R_{\rm ph}$). In the inner region above $R_{\rm eq}$, matters and photons are coupled up to the radius $R_{\rm ad}$. The photon-trapped radius ($R_{\rm ad}$) is defined by $t_{\rm diff} = t_{\rm dyn}$, where $t_{\rm diff}$ and $t_{\rm dyn}$ are the diffusion time and the dynamical time at $R_{\rm ad}$ \citep[see also][]{Piro2020ApJ}; 
\begin{equation}
    \frac{\tau_{\rm s}(R_{\rm ad})}{c} \frac{(R_{\rm out} - R_{\rm ad})R_{\rm ad}}{R_{\rm out}} = \frac{R_{\rm ad} - R_{\rm eq}}{v}.
\end{equation}
In addition, the colour radius ($R_{\rm c}$) is defined by $\tau_{\rm eff}(R_{\rm c}) \approx 1$. Then, the photospheric radius is given by $R_{\rm ph} = \max(R_{\rm ad}, R_{\rm c})$.

The temperature below the radius $R_{\rm ad}$ is decreasing adiabatically, following 
\begin{equation}
    T(r) = T_{\rm ad}\left( \frac{r}{R_{\rm ad}} \right)^{-2/3},
\end{equation}
where $T_{\rm ad}$ is the temperature at $R_{\rm ad}$ determined by adiabatic cooling, while the temperature above $R_{\rm ad}$ is determined by photon diffusion:
\begin{equation}
    T(r) = T_{\rm diff}\left( \frac{r}{R_{\rm ad}} \right)^{-3/4},
\end{equation}
where $T_{\rm diff}$ is the temperature at $R_{\rm ad}$ determined by diffusion. In the time-dependent outflow, the temperature at $R_{\rm ad}$ has a jump, i.e., $T_{\rm ad} \neq T_{\rm diff}$. To determine the temperature jump, we require an additional constraint given by $L_{\rm advec}(R_{\rm ad}) = L_{\rm diff}(R_{\rm ad})$, where $L_{\rm advec}(R_{\rm ad})$ is the advection luminosity, and $L_{\rm diff}(R_{\rm ad})$ is the diffusion luminosity \citep{Piro2020ApJ}. Then, 
\begin{equation}
    4\pi R_{\rm ad}^{2} a T_{\rm ad}^{4} \left( v - \frac{dR_{\rm ad}}{dt} \right) = - \frac{4\pi R_{\rm ad}^{2} ac}{3\kappa_{\rm s} \rho} \frac{\partial}{\partial r} T(r)^{4},
\end{equation}
where $a$ is the radiation constant, and $c$ is the light speed. Then, the temperature jump is determined as follows:
\begin{equation}
    \frac{T_{\rm ad}}{T_{\rm diff}} = \left\{ \frac{c}{v} \tau_{\rm s}(R_{\rm ad})^{-1} \left( 1 - \frac{1}{v}\frac{dR_{\rm ad}}{dt} \right)^{-1} \right\}^{1/4}.
\end{equation}
Note that, in the steady-state solution, the time-dependent term can be negligible, and then $T_{\rm ad}$ and $T_{\rm diff}$ can be connected without the temperature jump. 

Finally, assuming the blackbody radiation, the photospheric temperature ($T_{\rm ph}$) is determined as follows.
\begin{equation}
  T_{\rm ph} = \left( \frac{L}{4\pi R_{\rm ph}^{2} \sigma} \right)^{1/4},
\end{equation}
where $\sigma$ is the Stefan–Boltzmann constant, and $L$ is the luminosity determined by advection or diffusion; $L = L_{\rm advec} (R_{\rm ad})$ or $L = L_{\rm diff}(R_{\rm c})$.

\section{light-curve modelling and Application to FBOTs} \label{sec:3}

\subsection{General Properties of Time-Dependent Model} \label{sec:3.1}

\begin{figure*}
\includegraphics[scale=0.35]{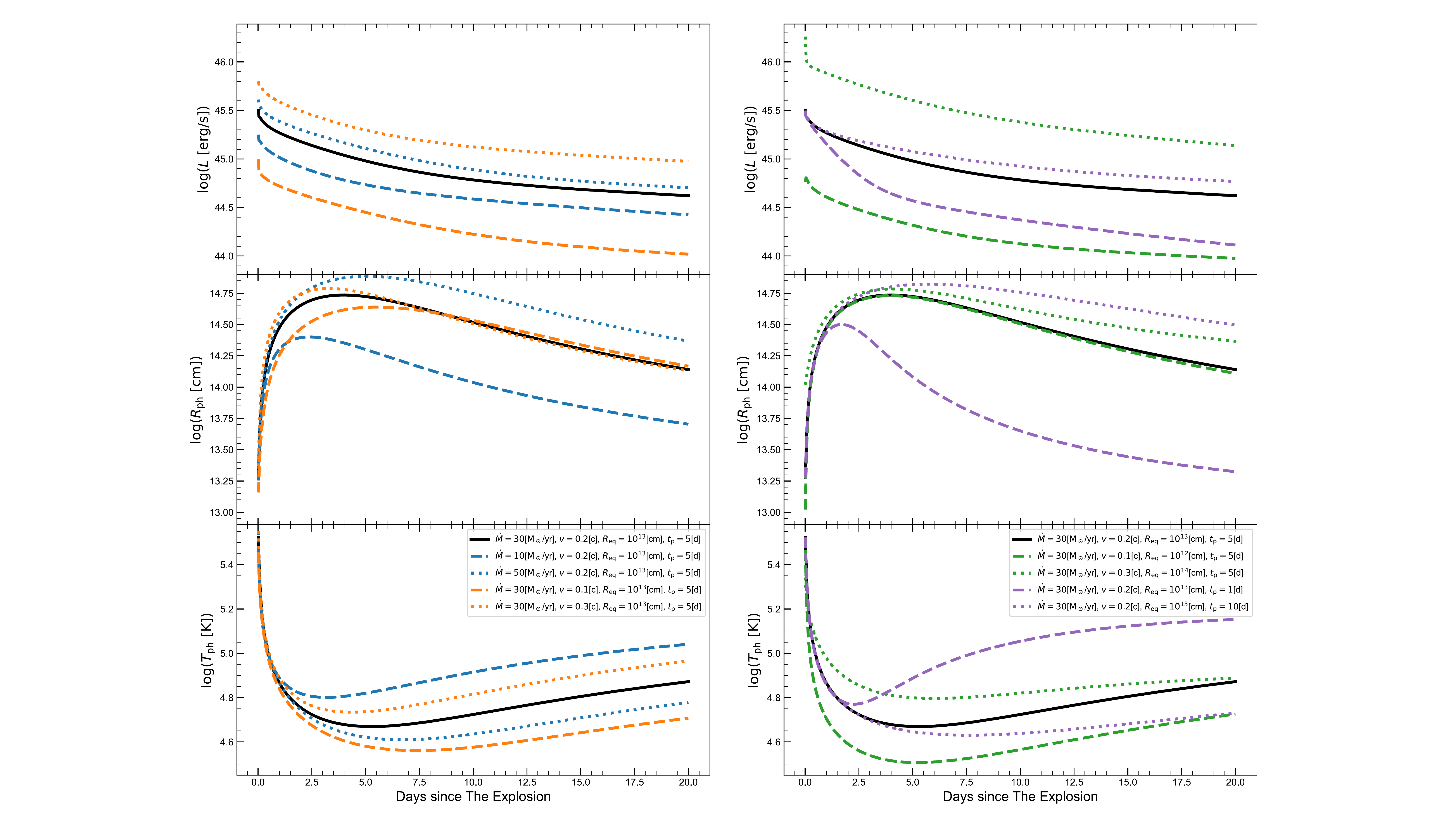}
\caption{The results of the light-curve modellings for some parameter sets, which are shown in the legends. We fix the power-law index as $\beta = -5/3$. The top, middle, and bottom panels show bolometric light curves, photospheric-radius evolution, and temperature evolution, respectively.}
\label{fig:1}
\end{figure*}

Figure \ref{fig:1} shows the results of the time-dependent wind-driven model for some parameter sets. The figure suggests that the peak luminosity is mainly determined by the outflow velocity, since the kinetic energy is proportional to $v^{2}$. Besides, Figure \ref{fig:1} shows that the rising timescale is determined by the outflow rate, while the decay rate is primarily determined by the outflow timescale ($t_{\rm 0}$) and the initial outflow rate  ($\dot{M}_{0}$). The outflow-launching radius ($R_{\rm eq}$) mainly impacts the temperature evolution; for smaller outflow-launching radii, the adiabatic cooling becomes more effective, and then the temperature at a given epoch will become lower.

\begin{figure}
\includegraphics[width=\columnwidth]{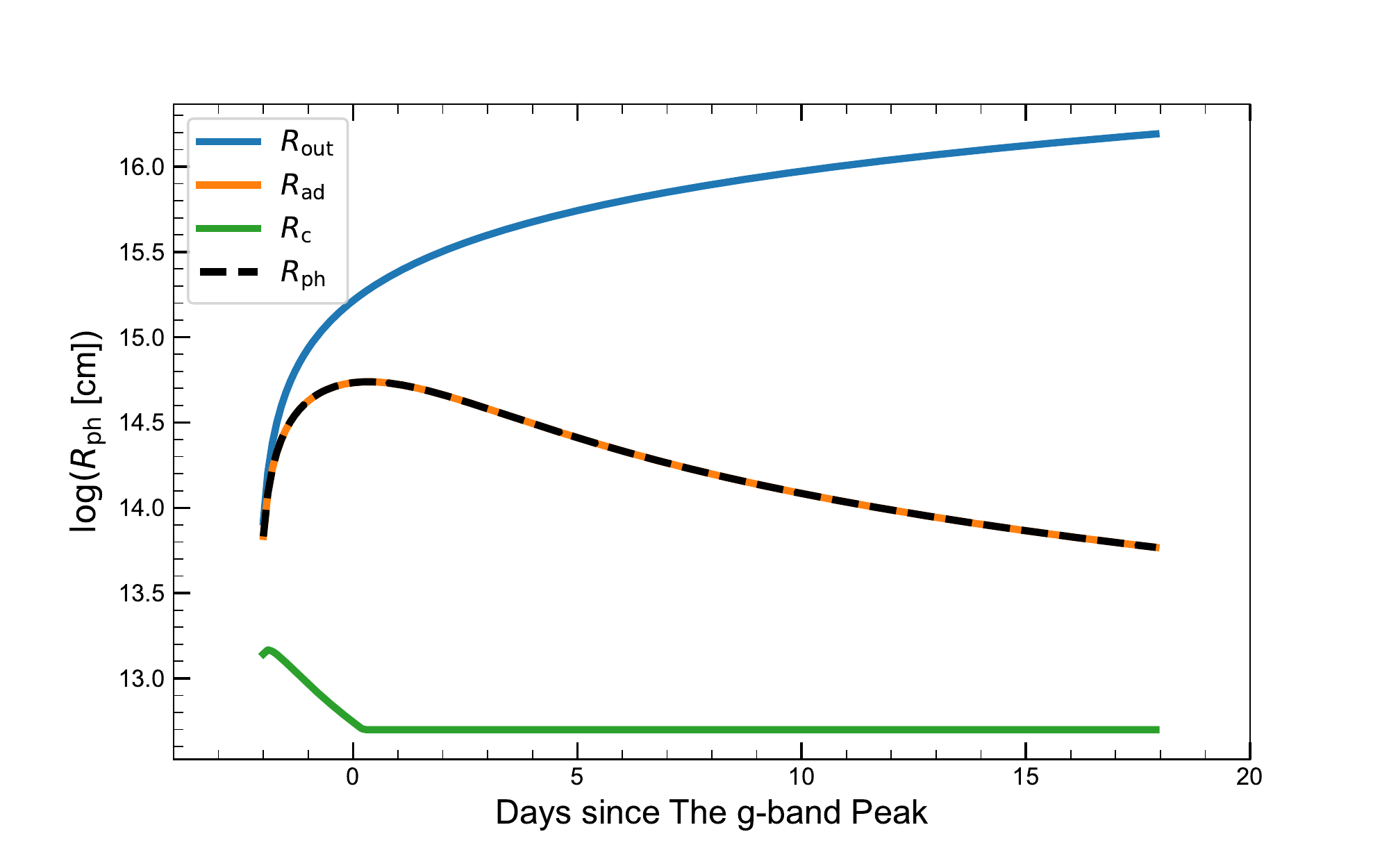}
\caption{The time evolution of the characteristic radius scales (see the legends), using the following parameter sets: $\dot{M}_{0}=30$\,M$_{\sun}$\,yr$^{-1}$, $v = 0.2c$, $t_{\rm 0} = 5$\,days, $R_{\rm eq} = 10^{13}$\,cm, and $\beta = -5/3$. 
}
\label{fig:2}
\end{figure}

In the initial phase, the outflow density and therefore the optical depth are so high that the outflow front is coupled with the photon-trapped radius (see also Figure \ref{fig:2}). Thus, the photospheric temperature rapidly declines immediately after the explosion due to adiabatic expansion. The peak temperature is $10^{5-6}$\,K, which indicates that FBOTs should show UV or X-ray flash similar to supernova shock breakouts.

Figure \ref{fig:2} plots the time evolution of the characteristic radius scales. It is seen that the photosphere follows the outflow front in the initial phase, while after a few days, the photosphere is decoupled from the outflow front as the optical depth decreases. This result indicates that the light-curve properties in the rising phase are determined by the initial outflow properties, i.e., the initial outflow rate and velocity. After the rising phase, the difference between the outflow front and photosphere increases, and then the system becomes represented approximately by the steady-state solution. 

\subsection{Application to FBOTs} \label{sec:3.2}

\begin{figure*}
\begin{subfigure}[]{0.5\linewidth}
\centering
\includegraphics[width=\columnwidth]{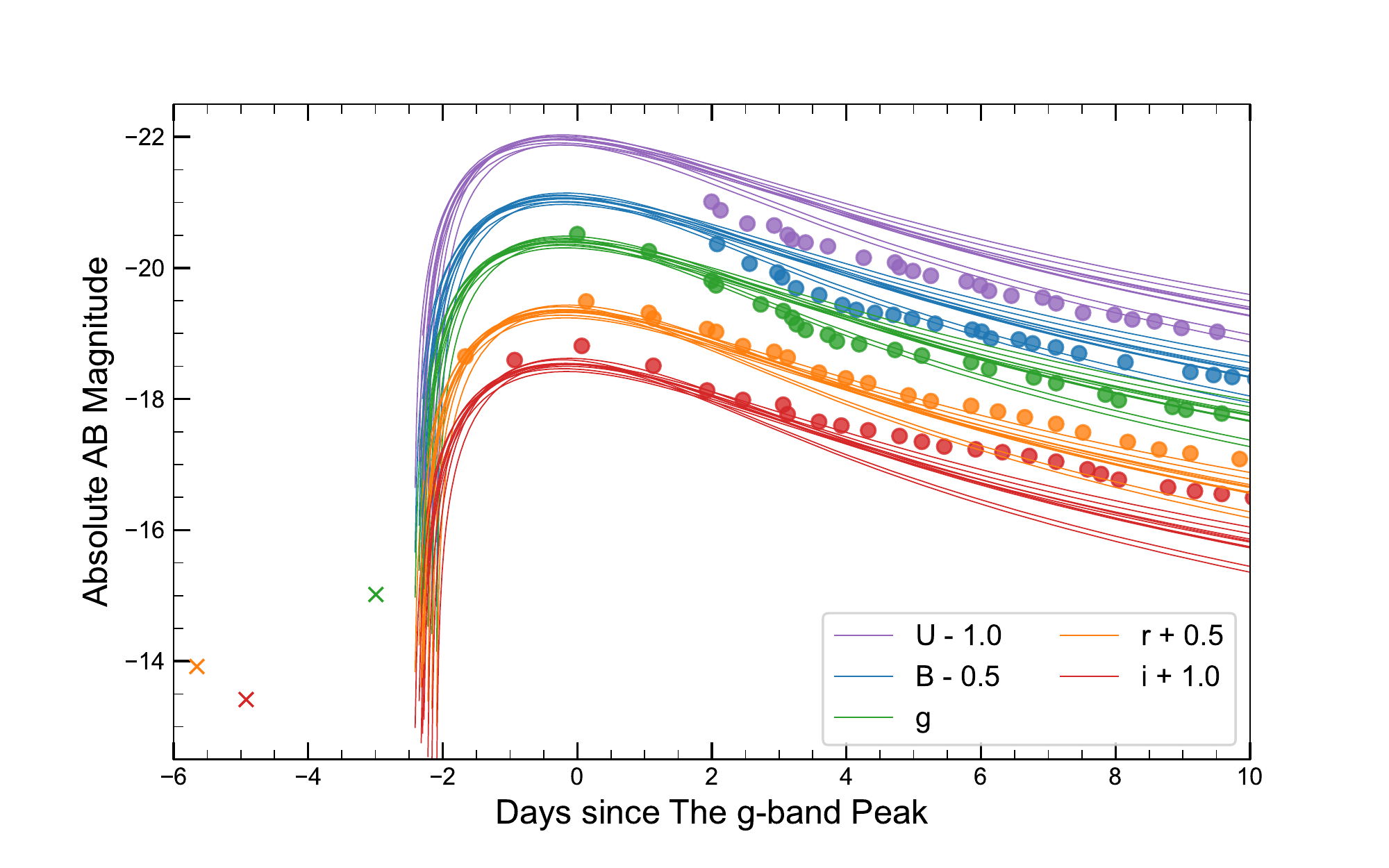}
\caption{AT2018cow}
\label{fig:lefttop}
\vspace{4ex}
\end{subfigure}
\begin{subfigure}[]{0.5\linewidth}
\centering
\includegraphics[width=\columnwidth]{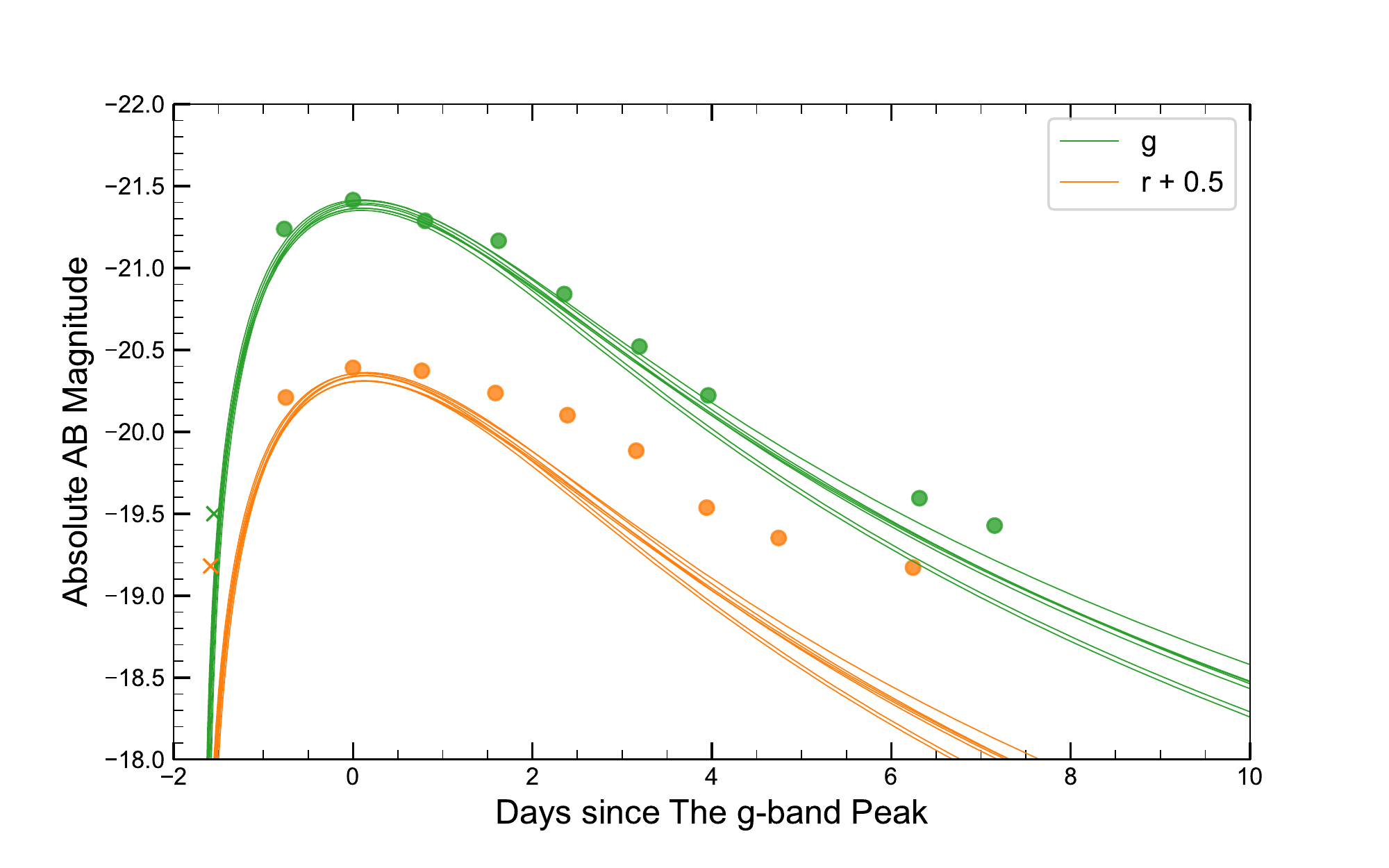}
\caption{AT2018lug}
\label{fig:righttop}
\vspace{4ex}
\end{subfigure}
\begin{subfigure}[]{0.5\linewidth}
\centering
\includegraphics[width=\columnwidth]{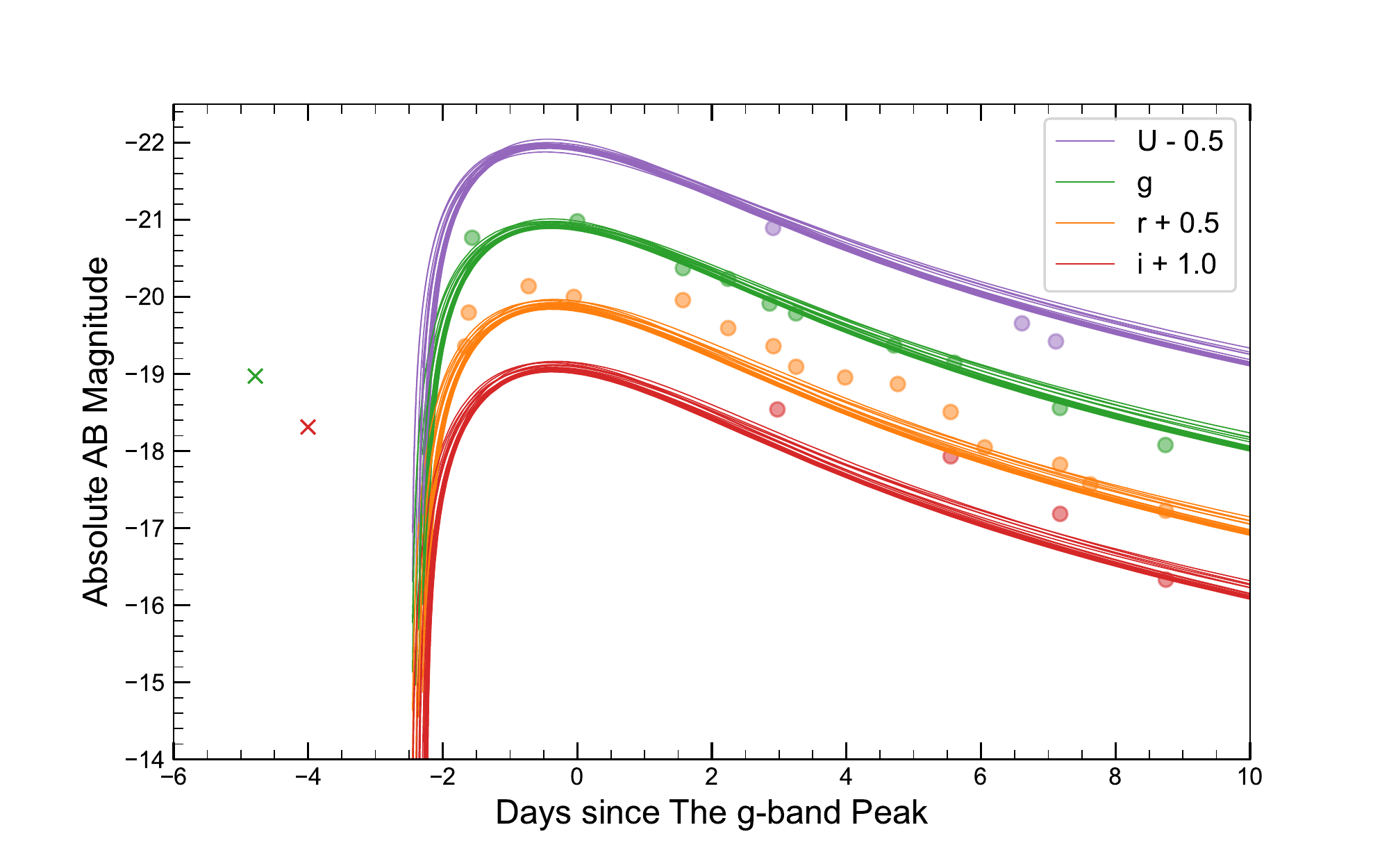}
\caption{AT2020mrf}
\label{fig:bottomleft}
\end{subfigure}
\begin{subfigure}[]{0.5\linewidth}
\centering
\includegraphics[width=\columnwidth]{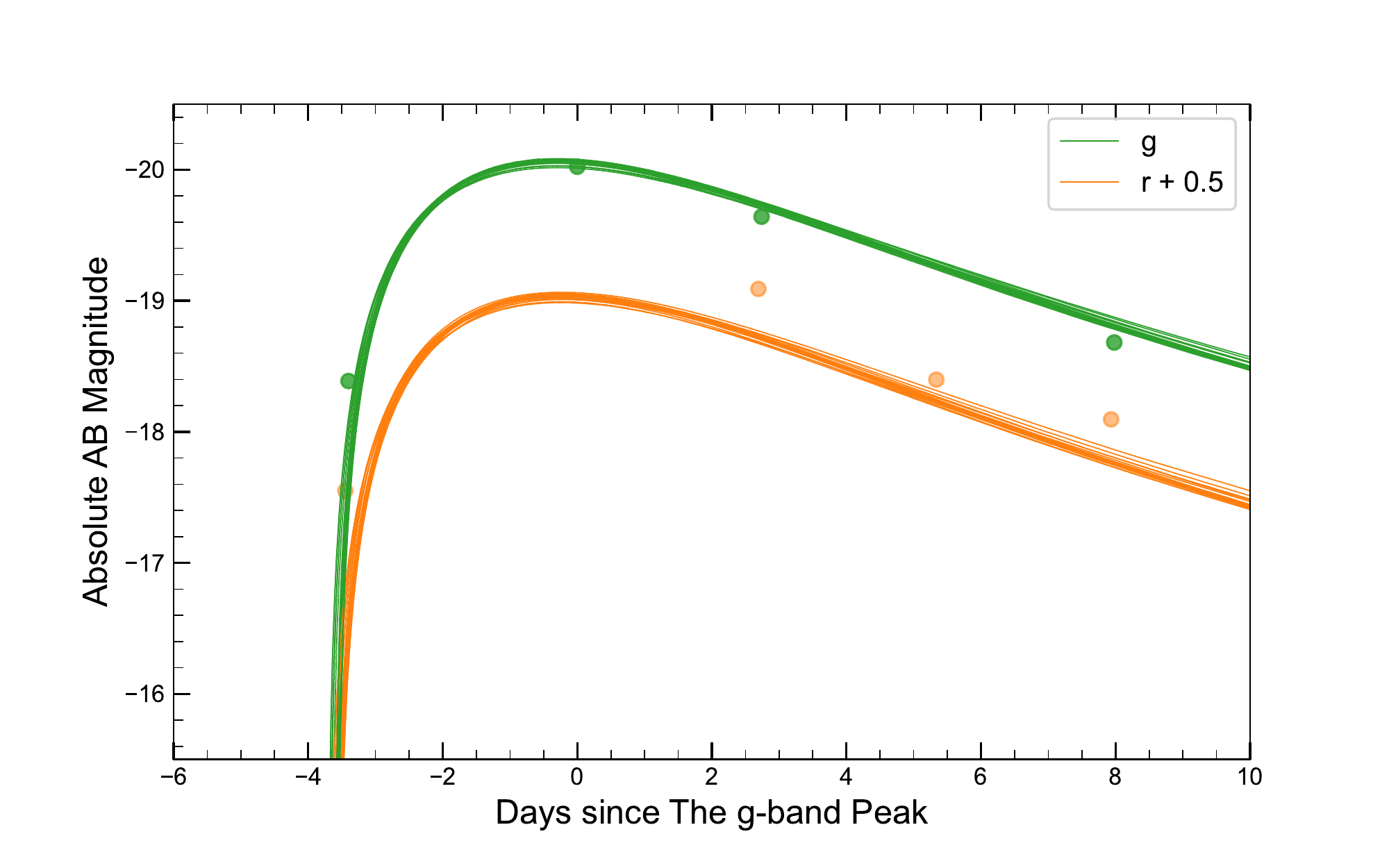}
\caption{AT2020xnd}
\label{fig:bottomright}
\end{subfigure}
\caption{
Early-phase multi-band light curves for a sample of FBOTs; (a) AT2018cow, (b) AT2018lug, (c) AT2020mrf, and (d) AT2020xnd, shown with the time-dependent wind-driven models with different parameter sets for different FBOTs. The solid lines in each panel show the best-fit light curves for each outflow-launching radius. For AT2018cow, AT2020mrf, and AT2020xnd, we computed 10 models with different outflow-launching radii, while for AT2018lug, we plot 7 models from $R_{\rm eq} = 4\times 10^{12}$\,cm to $1\times 10^{13}$\,cm, since the other 3 models did not have reasonable parameter sets.
}
\label{fig:3}
\end{figure*}

We apply the time-dependent model to a sample of well-observed FBOTs; AT2018cow, AT2018lug, AT2020mrf, and 2020xnd (see Figure \ref{fig:3}). We list the light-curve parameters, which are fitted by visual inspection, in Table \ref{tab:1}. \citet{Uno2020ApJ} proposed that AT2018cow is powered by mass accretion onto a central compact object, i.e., $\beta = -5/3$ \citep[see also][]{Piro2020ApJ}. Thus, in this work, we adopt the same power-law index as $-5/3$.

Unfortunately, lacking UV observations in the rising phase, the four model parameters; $\dot{M}_{0}$, $v$, $t_{0}$, and $R_{\rm eq}$, are degenerate. In particular, it is difficult to determine the outflow-launching radius, since $R_{\rm eq}$ affects the photospheric temperature, which is dominated by the UV emission (see also Figures \ref{fig:1} and \ref{fig:6}). Therefore, in this paper, we adopt $R_{\rm eq}$ as a free parameter; we compute the light curves with the outflow-launching radii from $R_{\rm eq} = 1\times 10^{12}$\,cm to $R_{\rm eq} = 1\times 10^{13}$\,cm with an increment of $1\times 10^{12}$\,cm. Then, we estimated the acceptable range of the other three parameters based on the wind-launched radius space (see Table \ref{tab:1}), and evaluated their uncertainties. Note that we plot the well-fitted 10 light curves in Figure \ref{fig:1}, except for AT2018lug. For AT2018lug, we cannot fit the light curves with $R_{\rm eq} = 1-3\times 10^{12}$\,cm due to the strict detection upper limit, and then we plot only 7 light curves and the uncertainty of other parameters are evaluated from the 7 models.

Figure \ref{fig:3} shows multi-band light curves of FBOTs. We construct the multi-band light curves assuming blackbody emission with the model photospheric temperature as convolved with the filter functions. To explain the fast-rising light curve and high peak luminosity of FBOTs, high outflow rates ($\dot{M}_{0} \gtrsim 30$\,M$_{\sun}$\,yr$^{-1}$) and fast outflow velocities ($v\gtrsim 0.2c$) are required. The energetic outflow supports the idea that the central engine may be related to a relativistic system, e.g., a BH. Besides, the rapid decline rate suggests that the outflow rate evolves in a timescale of a few days, which indicates that the typical dynamical timescale of the outflow related to the central object is also a few days, i.e., a stellar-/intermediate-mass BH.

The models shown in Figure \ref{fig:3} can explain the light-curve properties in the rising phase and post-peak phase until $\lesssim 5$\,days. In the post-peak phase, the gaps between the observed magnitudes and models become substantial; this is mainly because we assume a constant outflow velocity, whereas the outflow velocity is inferred to be monotonically decreasing in the observation \citep[e.g.,][]{Perley2019MNRAS}. 

\begin{figure}
\includegraphics[width=\columnwidth]{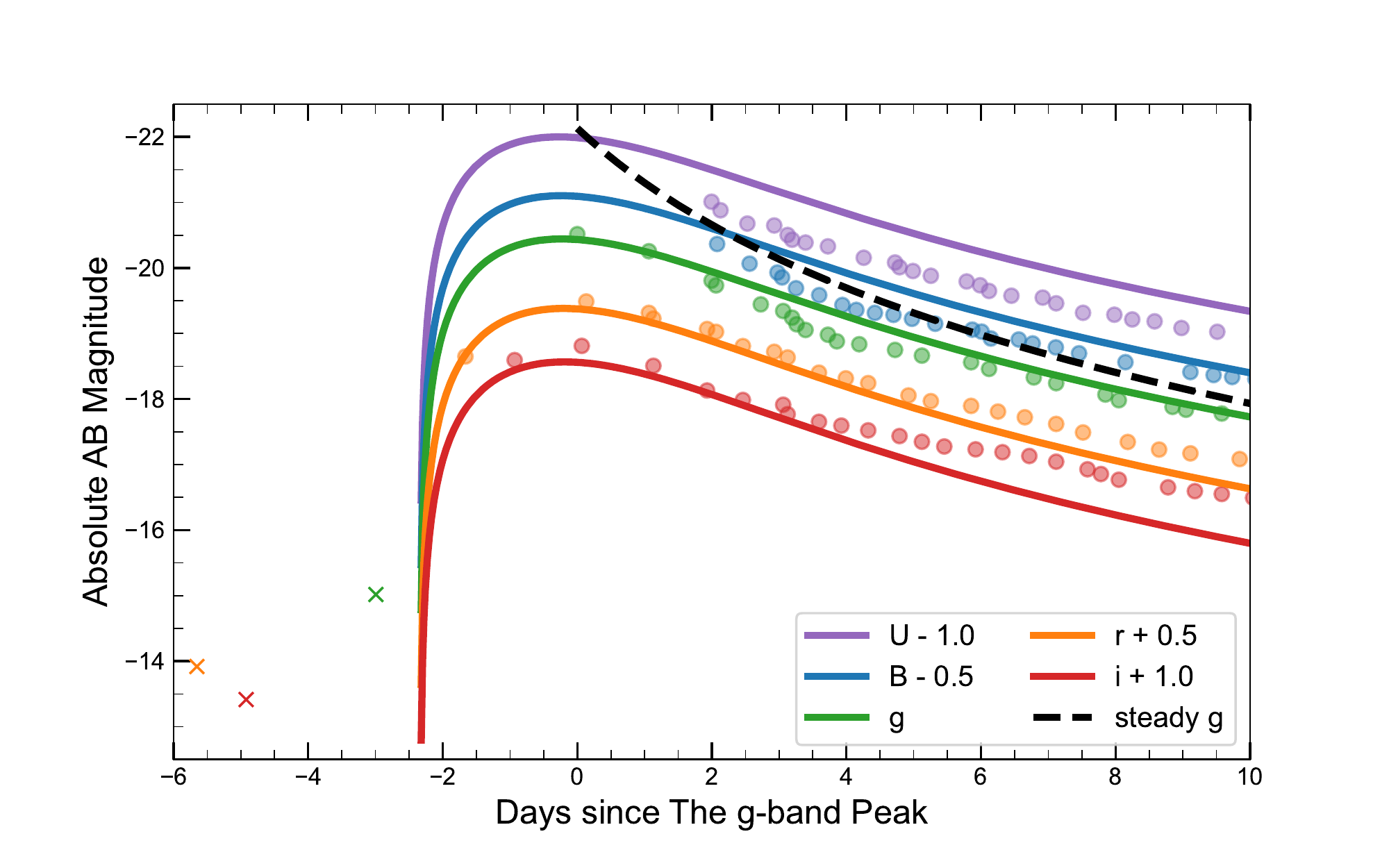}
\caption{
Multi-band light-curve model of AT2018cow with the mean light-curve parameters. We adopt $R_{\rm eq}$ as $5\times 10^{12}$\,cm. We also plot the $g$-band light curve computed with the steady-state model with the same parameter set as the corresponding time-dependent model. The steady-state model starts following the time-dependent model a few days after the peak, which indicates that the steady-state model can apply to the post-peak light curves.
}

\label{fig:4}
\end{figure}

In the initial phase, the light-curve behaviours are essentially determined by the outflow front, i.e., by the properties of the outflow ejected at the very beginning. However, in the post-peak phase, the photosphere recedes inward, and then the time dependence of the outflow properties starts affecting the light-curves behaviours. Therefore, to model the post-peak phase, one has to take into account the possible evolution of all the outflow properties; the present time-dependent formalism is limited by the assumption of the constant velocity. Instead, it can be (and has been) easily taken into account in the steady-state model \citep{Uno2020ApJ} that can thus be readily applicable to the post-peak phase. Indeed, if we take the same parameters (and their evolution) for the time-dependent and steady-state models, the predicted light-curve properties marge in the post-peak phase as shown in Figure \ref{fig:4}; the time-dependent model follows the steady-state model at a few days since the peak and thereafter, and thus one can apply the steady-state solution to the post-peak phase.

\begin{table}
\begin{center}
\caption{The light-curve parameters for a sample of FBOTs }
\label{tab:1}
\begin{tabular}{lccccc}
\hline
Object & $\dot{M}_{0}$ & $v$ & $t_{0}$ & $M_{\rm tot}$  & $E_{\rm tot}$ \\
& [M$_{\sun}$\,yr$^{-1}$] & [c] & [days] & [M$_{\sun}$] & [$10^{52}$\,erg] \\
\hline
      18cow & $37 \pm 7^{*}$ & $0.31 \pm 0.02$ & $2.4 \pm 0.4$ & $0.19 \pm 0.02$ & $1.7 \pm 0.4$\\
      18lug & $40 \pm 3$ & $0.38 \pm 0.01$ & $2.0 \pm 0.2$ & $0.18 \pm 0.01$ & $2.3 \pm 0.1$\\
      20mrf & $26 \pm 2$ & $0.18 \pm 0.02$ & $5.2 \pm 0.2$ & $0.20 \pm 0.02$ & $0.6 \pm 0.2$\\
      20xnd & $37 \pm 5$ & $0.31 \pm 0.02$ & $2.2 \pm 0.1$ & $0.18 \pm 0.02$ & $1.5 \pm 0.3$\\
\hline
\end{tabular}
\end{center}
$^{*}$: the uncertainty of each parameter is 1-$\sigma$ statistical uncertainties evaluated from all well-fitted models for each object. \\
\end{table}

In Table \ref{tab:1}, we show the cumulative ejecta masses ($M_{\rm tot}$) and kinetic energies ($E_{\rm tot}$) up to 5 days after the explosion, for a sample of FBOTs, as derived by the present models. The typical ejecta mass and energy budget are $\sim 0.2$\,M$_{\sun}$ and $\sim 10^{52}$\,erg, respectively. Note that the kinetic energy here can be regarded as an upper limit because we here assume a constant velocity, while the observations indicate that their velocities decrease monotonically. In addition to the power-law decay in the outflow rate, the mass and energy budgets also suggest that FBOTs are powered by gravitational energy release around central BH, e.g., BH-forming SNe or TDEs \citep[e.g.,][]{Uno2020ApJ}. The scenario is also supported by the outflow radii of $10^{12-13}$\,cm, which are typical radius scales of supergiants \citep{Uno2020ApJ} or TDEs \citep[the disk-wind-launched radius or self-interaction radius;][]{Uno2020ApJL}.

\section{Discussion} \label{sec:4}

\subsection{Future Observations}\label{sec:4.1}

\begin{figure*}
\begin{subfigure}[]{0.5\linewidth}
\centering
\includegraphics[width=\columnwidth]{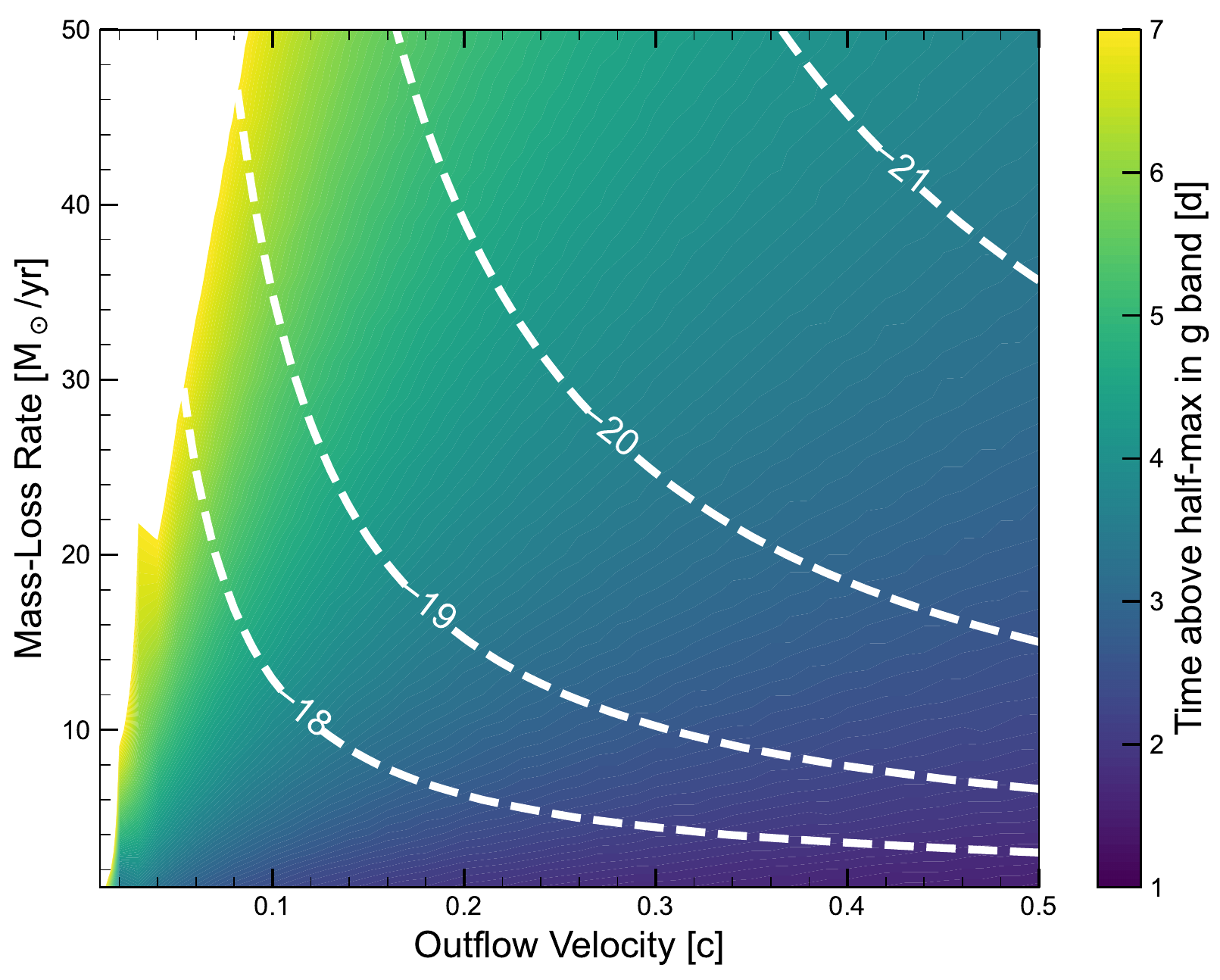}
\caption{ZTF/$g$ band}
\label{fig:lefttop}
\end{subfigure}
\begin{subfigure}[]{0.5\linewidth}
\centering
\includegraphics[width=\columnwidth]{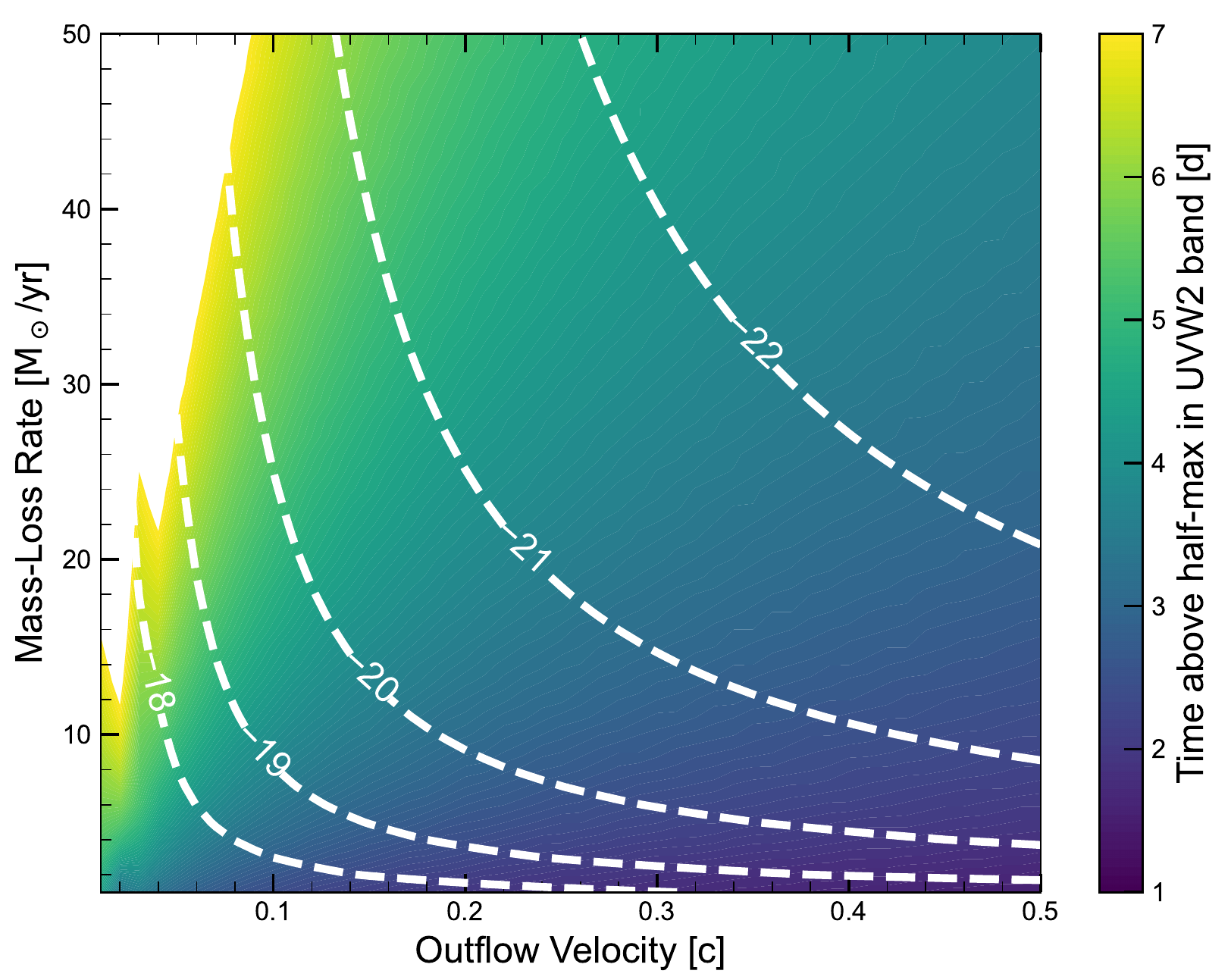}
\caption{Swift/$UVW2$ band}
\label{fig:bottomleft}
\end{subfigure}
\caption{The typical timescale and peak absolute magnitude ((A): ZTF/$g$ band and (B): Swift/$UVW2$ band), as functions of the outflow rate and velocity. The colour contour shows the typical timescale, which is the time above half-maximum brightness (for the contour scale, see the colour bars on the right side of each panel). The white contour shows the peak magnitude (for the contour scale, see the labels in each panel.)}
\label{fig:5}
\end{figure*}

According to the time evolution of the estimated photospheric temperature (see Figure \ref{fig:1}), we suggest an important role of X-ray/UV-triggered observation with space telescopes and rapid follow-up observations with ground-based telescopes, to go beyond the present situation where FBOTs have been discovered by optical surveys, such as ZTF or ATLAS. We compute the magnitudes and timescales with the time-dependent wind-driven model for various parameter sets (Figure \ref{fig:5}). The peak magnitudes in the UV bands are brighter than in the optical bands, and the duration of the UV emission is a bit longer. Figure \ref{fig:5} supports that UV surveys may have some advantages to discover FBOTs. Moreover, FBOTs tend to be discovered at high redshifts due to their rarity. Considering the high redshifts, the bright UV emission in the rest frame may benefit optical observations in the observed frame.

Unfortunately, the field of view of currently operating UV telescopes, e.g., Swift/UVOT \citep{Roming2005SSR} and ASTROSAT \citep{Singh2014SPIE}, is not sufficiently wide, and thus they are not suitable for surveying FBOTs. Wide-field UV telescopes are necessary to capture the initial, rising light curves of FBOTs and to elucidate the enigmatic nature of FBOTs. The present results suggest that it is essential to perform survey observations with high cadence of less than one day, and to establish a system to operate ground-based telescopes immediately after the discovery.

In addition, Figure \ref{fig:5} allows us to easily link observational properties of rapidly-evolving transients to physical properties, in the framework of the wind-driven model. It will thus be useful not only in planning future observations but also in interpreting observational data.

\subsection{Constraints on The Wind-Driven Model from UV Observations}\label{sec:4.2}

\begin{figure}
\includegraphics[width=\columnwidth]{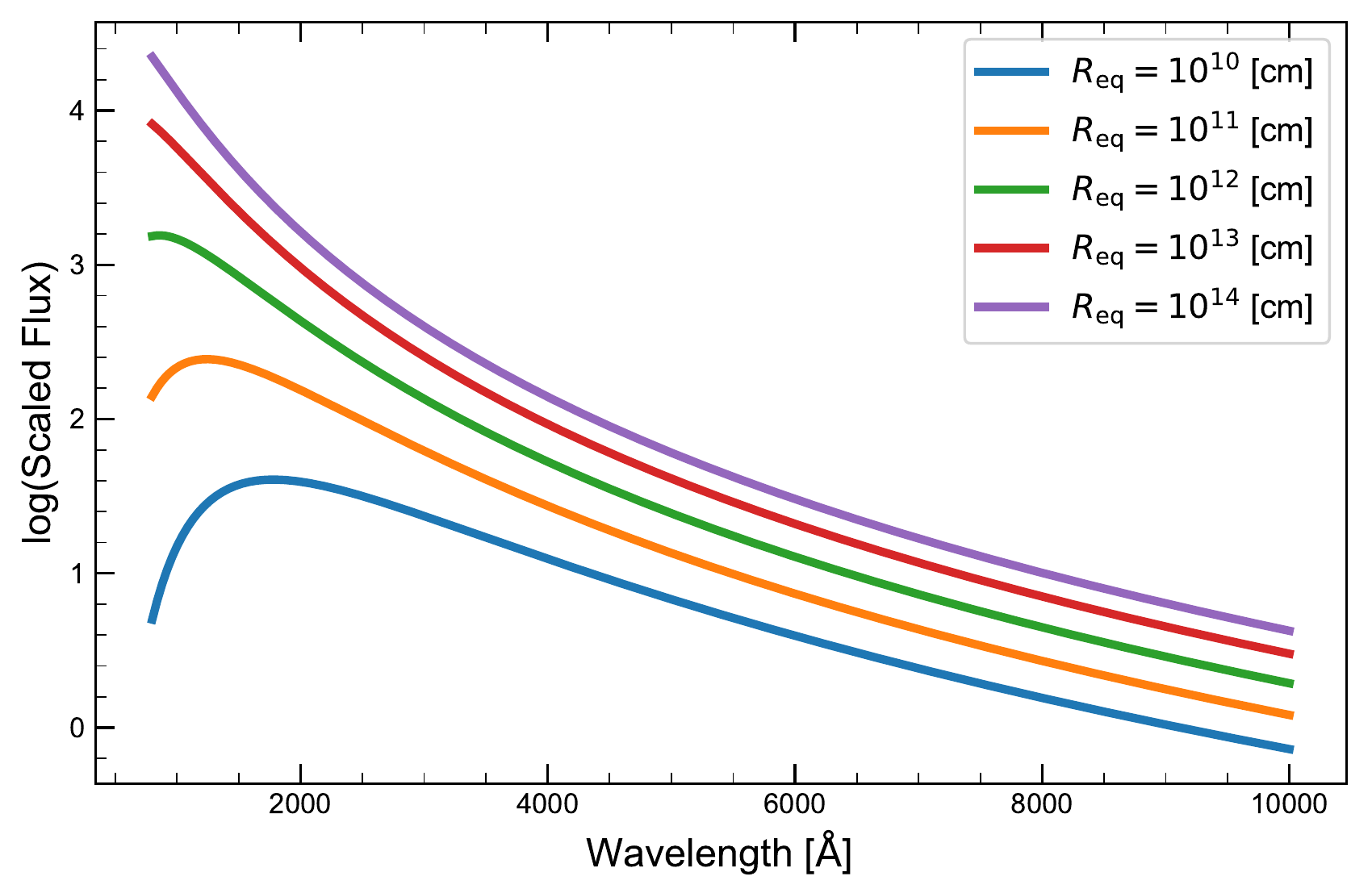}
\caption{Spectral energy distributions of some models at the $g$-band peak phase, adopting different outflow-launching radii. For a smaller outflow-launching radius, the SED peaks become closer to the optical region because adiabatic cooling becomes more effective.}
\label{fig:6}
\end{figure}

UV observations also provide us with additional constraints on our model. The typical temperature of FBOTs is estimated to be $\gtrsim 30000$\,K, and thus the spectral energy distribution (SED) peak is located in the UV region while the optical emission is dominated by the Rayleigh-Jeans tail (see also Figure \ref{fig:6}). Therefore, to determine the photospheric temperature, we need UV observations. 

In our model, the outflow-launching radius is a key physical parameter that mainly determines the photospheric temperature (see also Section \ref{sec:3.1}). Figure \ref{fig:6} shows scaled SEDs at the $g$-band peak phase, for models with different outflow-launching radii. It shows that a small outflow-launching radius predicts low photospheric temperature. Therefore, having a good estimate of the photospheric temperature is essential to obtain a unique solution in the present model framework. 

This is indeed a critical limitation in the present situation. For example, the well-observed FBOT sample has little information on UV behaviours in the rising phase, and thus it is difficult to remove the parameter degeneracy and determine the outflow-launching radius. UV observations will lead to resolving the degeneracy of the model parameters, playing an important role in characterizing the physical properties of individual FBOTs. Recently, a new FBOT candidate with the initial UV observation available has been discovered \citep[MUSSES2020J;][]{Jiang2022ApJ}. We plan to present detailed models for this object in a forthcoming paper, to demonstrate how the parameter degeneracy can be solved once the UV information is available.

\section{Conclusions} \label{sec:5}

In the present paper, we have proposed a time-dependent wind-driven model. The rising timescale is determined by the initial outflow rate, while the post-peak decay rate is mainly determined by the outflow timescale. Besides, the peak magnitude is determined by the (initial) outflow velocity. The initial temperature is estimated to be $10^{5-6}$\,K, which suggests that FBOTs will show UV or X-ray flash similar to supernova shock breakouts.

We apply the model to a sample of well-observed FBOTs; AT2018cow, AT2018lug, AT2020mrf, and AT2020xnd. They require high outflow rates ($\sim 30$\,M$_{\sun}$\,yr$^{-1}$) and fast velocity ($\sim 0.2-0.3c$), and then the typical ejecta mass and energy budgets are $\sim 0.2$\,M$_{\sun}$ and $\sim 10^{52}$\,erg, respectively. The energetic outflow, as derived by applying the time-dependent model in the rising properties, supports that the central engine of FBOTs may be related to a BH; this is consistent with the results based on the `steady-state' model as applied to the post-peak properties \citep{Uno2020ApJ}.

We have discussed the advantages of UV observations in discovering FBOTs and characterizing their natures. In our formalism, the peak magnitudes in the UV bands are brighter, and the timescale of the light curves in the UV bands is a bit longer than in optical wavelengths. We suggest that UV telescopes with a wide field of view should play a key role in discovering FBOTs in future observations.

\section*{Acknowledgements}

The authors thank Ji-an Jiang for valuable discussion. The authors also acknowledge `1st Finland-Japan bilateral meeting on extragalactic transients' (partly supported by the JSPS Open Partnership Bilateral Joint Research Projects between Japan and Finland; JPJSBP120229923), which gave us a good opportunity to discuss this model. K.U. acknowledges financial support from Grant-in-Aid for the Japan Society for the Promotion of Science (JSPS) Fellows (22J22705). K.U. also acknowledges financial support from AY2022 DoGS Overseas Travel Support, Kyoto University. K.M. acknowledges support from the JSPS KAKENHI grant JP18H05223, JP20H00174, and JP20H04737.

\section*{Data Availability}
No new observational data were analysed in this research.
The model results will be available on request.

\bibliographystyle{mnras}
\bibliography{manuscript} 

\bsp	
\label{lastpage}
\end{document}